\documentclass[conference, 9pt]{IEEEtran}
\IEEEoverridecommandlockouts
\usepackage{cite}
\usepackage{amsmath,amssymb,amsfonts}
\usepackage{algorithmic}
\usepackage{multirow}
\usepackage{graphicx}
\usepackage{textcomp}
\usepackage{xcolor}
\usepackage{fancyhdr}
\usepackage[hyphens]{url}

\usepackage{comment}
\usepackage{epstopdf}
\usepackage{amssymb}
\usepackage{tikz}
\usepackage{booktabs,caption,fixltx2e}
\usepackage[flushleft]{threeparttable}
\usepackage{dblfloatfix}

\usepackage{caption}
\usepackage{xcolor}
\definecolor{myBlue}{rgb}{0,0.0,1.0}

\usepackage{xspace}

\def\BibTeX{{\rm B\kern-.05em{\sc i\kern-.025em b}\kern-.08em
    T\kern-.1667em\lower.7ex\hbox{E}\kern-.125emX}}

\fancypagestyle{firstpage}{
  \fancyhf{}

  \fancyhead[C]{\normalsize{Appears in the 40th IEEE International Conference on Computer Design (ICCD), 2022
    }}
  \fancyfoot[C]{\thepage}
}

\title{LightNorm: Area and Energy-Efficient Batch Normalization Hardware for On-Device DNN Training
\thanks{This work was supported by Samsung Research Funding Incubation Center
of Samsung Electronics under Project Number SRFC-IT1902-03, and
the EDA tool was supported by the IC Design Education Center (IDEC) in South Korea.

J. Kung is the corresponding author (jhkung@dgist.ac.kr).
}}
\author{
\begin{large}Seock-Hwan Noh, Junsang Park, Dahoon Park, Jahyun Koo, Jeik Choi, Jaeha Kung
\end{large} \\ 
\begin{small} Dept. of Electrical Engineering and Computer Science, DGIST, Daegu, South Korea
\end{small} \\
\begin{small} 
\{nosh3332, jshparksh, pdh930105, jhkoo, ikchoi, jhkung\}@dgist.ac.kr
\end{small}
}
\begin{document}
\maketitle
\thispagestyle{firstpage}
\pagestyle{plain}

%%%%%%%%% -- need to check -- %%%%%%%%%%
% 1. scale factor
% 2. MB2 accuracy (Lowr than MB1, but it is wrong)
% 3. Remove the figure 3.
% 4. Need to change fig.8 (FP16 -> FP8+8)
% 5. Only FP8 and FP32...?
% 6. Blocking with channel direction and fmap direction. How are they different?/
% 7. Figure out a dynamic range of range BN

%%%%%% -- PAPER CONTENT STARTS-- %%%%%%%%

\begin{abstract}

When training early-stage deep neural networks (DNNs), generating intermediate features
via convolution or linear layers occupied most of the execution time.
Accordingly, extensive research has been done to reduce the computational burden of the convolution or linear layers.
In recent mobile-friendly DNNs, however, the relative number of operations involved in processing these layers 
has significantly reduced.
As a result, the proportion of the execution time of other layers, such as batch normalization layers, has increased.
Thus, in this work, we conduct a detailed analysis of the batch normalization layer to efficiently reduce the runtime overhead in the batch normalization process.
Backed up by the thorough analysis, we present an extremely efficient batch normalization, named LightNorm, and its associated hardware module.
In more detail, we fuse three approximation techniques that are i) low bit-precision, ii) range batch normalization, and iii) block floating point.
All these approximate techniques are carefully utilized not only to maintain the statistics of intermediate feature maps, but also to minimize the off-chip memory accesses.
By using the proposed LightNorm hardware, we can achieve significant area and energy savings during the DNN training without hurting the training accuracy. This makes the proposed hardware a great candidate for the on-device training.
\end{abstract}

\section{Introduction}\label{sec:introduction}
Recently, deep learning has been applied in many fields of our daily life such as autonomous driving, computer vision and language modeling~\cite{driving, gpt3, blur}. 
Prior to the deployment of deep learning in these applications, training weight parameters of deep neural networks (DNNs) should be preceded. 
However, training DNNs is a computationally expensive task due to the large amount of weight parameters and thousands of training iterations on large datasets. 
It requires several days to train the large DNN models even if the training is done on state-of-the-art GPUs or custom NPUs. 
For instance, training ResNet-50 takes 29 hours on 8 Tesla P100 GPUs~\cite{resnet} and training BERT requires 16 TPU-v3 chips for 3 days~\cite{bert}.
This prohibits the training at end devices thus the training is mostly done at much powerful cloud servers.

% Low bit-precisions and examples
Generally, training DNN models is performed with a IEEE single precision format, i.e., FP32. 
However, training DNNs with the FP32 format on a large dataset accompanies excessive hardware costs in terms of energy consumption, memory capacity and communication cost. 
As a remedy for this challenge, reducing bit-width has been actively explored to effectively minimize the hardware cost. 
For instance, M. Paulius et al. have proposed a mixed-precision training method for the recent GPU to improve the training throughput~\cite{mix_training}. 
With the mixed-precision training, multiplications are performed in FP16 while accumulations are performed in FP32.
Unconventional data representations suited at DNN training, such as bfloat16~\cite{bfloat16, nnp-t} and block floating point (BFP) representation~\cite{hybrid_bfp, flexblock}, have been studied as well.
To enable DNN training at much lower hardware cost, possibly at the edge, researchers have explored low-precision training using FP8 (i.e., 8-bit floating point) with the support of squeeze and shift operations~\cite{parameter8_training} or exponent biases~\cite{minifloat} to cover a wide dynamic range of the original data distribution.
However, the prior work only allow low-precision operations on DNN layers, i.e., convolution (Conv) or fully-connected (FC) layers, leaving non-DNN layers including batch normalization to be processed with FP32.

\begin{figure}[t]
    \centering
    \includegraphics[scale=0.48]{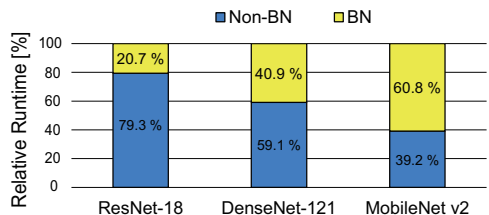}
    \caption{Runtime breakdown of BN and non-BN layers per training iteration tested on three CNN benchmarks. The models are trained with the recent GPU (i.e., NVIDIA RTX 3090) on ImageNet dataset. `Non-BN' represents the training time consumed by all operations except BN layers. They are Conv/FC layers, activation functions, and pooling layers.}
    \label{fig:exeution_time}
\end{figure}

% BN is also important.
However, as the DNN model evolves over time, the relative execution time of the previously time-consuming Conv/FC layers decreases. 
For example, early convolutional neural networks (CNNs), such as AlexNet and VGGNet, consume most of the training time (83.9$\sim$95.1\%) at Conv or FC layers~\cite{bn_sysml}.
Fig.~\ref{fig:exeution_time} shows the relative execution time between the batch normalization (BN) layers and non-BN layers in three representative CNN benchmarks.
Here, Conv/FC layers are included in the non-BN layers.
Since the design of recent CNN models focuses on reducing the computational cost of Conv/FC layers, e.g., by reducing the kernel size~\cite{resnet} or introducing a new layer structure~\cite{densenet, mobilenet_v2}, the relative execution time of non-BN layers significantly reduces on mobile-friendly CNNs.
As shown in Fig.~\ref{fig:exeution_time}, only 39.2\% of the total runtime is consumed by the non-BN layers for MobileNetV2 while 60.8\% of the runtime is consumed by the BN layers.
In other words, for training mobile-friendly CNN models, the runtime overhead of the overlooked BN layers becomes significant.
Therefore, to enable on-device DNN training with limited hardware resources, we should explore reducing the complexity of not only the Conv/FC layers, but also the BN layers.
Recently, a couple of research works have been presented that try to apply reduced precision on the BN layers, e.g., FP16~\cite{norm_matter} or bfloat16~\cite{bfloat16}.
However, these previous works reduce the bit-width of data to store them in a smaller off-chip memory and to minimize the energy consumption of data accesses, while still limiting the computation to 16-bit multiplications with FP32 additions.

In this paper, we present an extremely compute- and memory-efficient BN process, named LightNorm, by fusing three approximation techniques, which are i) low bit-precision, ii) range batch normalization (RN), and iii) block floating point (BFP).
By using RN, the required dynamic range in representing the intermediate values during the BN process is constrained, which allows lower precision to work well on the BN layers.
In addition, converting to a BFP representation prior to storing feature maps to DRAM helps minimize the communication cost incurred by expensive DRAM accesses.
Then, we present a customized hardware for LightNorm which realizes area and energy-efficient BN hardware for on-device DNN training.
By taking benefits of all these approaches, we were able to design an extremely lightweight BN hardware, which takes up 16.2$\times$ smaller area and consumes 15.4$\times$ lower power compared to the previous BN hardware designs.

The main contributions of our work are as follows:
\begin{enumerate}
    \item \textbf{Hardware-friendly BN:} We explored the possible combinations of approximation techniques to design a low-complexity BN layer, named LightNorm, possibly realizing on-device training for recent mobile-friendly DNN models.
    We first utilize range batch normalization to push the quantization level down to FP10.
    Then, to further reduce the energy consumption, we group tensors so that exponents are shared.
    \item \textbf{Hardware Implementation:}  We designed LightNorm hardware and compared its efficiency to other BN methods. Moreover, we designed a training accelerator equipped with the LightNorm hardware to evaluate the system-level energy efficiency.
    Compared to training accelerators with different precision levels and BN types, the area reduces by 1.2$\sim$4.1$\times$ and the energy consumption decreases by 1.3$\sim$5.0$\times$.
\end{enumerate}

\begin{comment}
The remainder of the paper is organized as follows:
Section~\ref{sec:background} presents the required background on the BFP-based DNN training; 
Section~\ref{sec:motivation} discusses limitations of previously proposed
precision-scalable MAC arrays when the training is considered;
Section~\ref{sec:reconfig} describes how various precision levels and layer types are supported by FlexBlock; 
Section~\ref{sec:overall_arch} explains the overall architecture; 
Section~\ref{sec:methodology} describes a software framework and hardware implementation that are used to validate 
the effectiveness of FlexBlock; Section~\ref{sec:experiments} discusses experimental results; 
Section~\ref{sec:related} presents the related work on DNN training accelerators, 
and concludes in Section~\ref{sec:conclusion}.
\end{comment}

\section{Backgrounds}\label{sec:background}
\subsection{Batch Normalization}\label{sec:BN}
A batch normalization (BN) is an essential process in deep learning to adjust input distribution to allow faster convergence and improved training accuracy. 
The concept of the BN was first introduced in~\cite{batchnorm} to address the problem of internal covariate shift. 
For a mini-batch of input tensor per channel, i.e., $X^{(c)} \in \mathbb{R}^{B\times H \times W}$, the BN is done by
\begin{equation}\label{eq:bn}
    y_i = \gamma \cdot {\frac{x_i - \mu}{\sqrt{Var[X^{(c)}] + \epsilon}}} + \beta, 
\end{equation}
where $c$ is the channel index, $x_i$ is the $i^{th}$ element in $X^{(c)}$, $B$ is the mini-batch size, $H$ and $W$ are the height and width of a feature map, respectively, $\mu$ is the expectation over $X^{(c)}$, 
%$Var[X]$ is an expectation of $(X-\mu)^2$, 
$Var[X^{(c)}]$ is the variance of $X^{(c)}$, and $\epsilon$ is the value to prevent the denominator from being zero.
The $\gamma$ and $\beta$ are trainable parameters for each channel `$c$' that improve the training accuracy.
Due to the importance of the BN process for training a better DNN model, it is a common design strategy to add BN layers in the DNN model~\cite{batchnorm_mit, understanding_batchnorm, batchnorm_residual_block, mit_book}.

%\subsection{Range Normalization (RN)}\label{sec:RN}
However, the computation of a conventional BN process in Eq.~(\ref{eq:bn}) requires complex arithmetic functions, i.e., square root and division,
that incur considerable hardware overhead when designing a DNN accelerator~\cite{range_bn}.
To reduce the complexity of realizing the conventional BN computations, a range batch normalization (RN) has been proposed~\cite{range_bn}.
The main idea of the RN is to replace the denominator term in Eq.~(\ref{eq:bn}) to a range of input distribution. 
The RN is performed by
\begin{equation}\label{eq:range_norm}
    \hat{y}_i = \gamma\cdot\frac{x_i - \mu}{C(B) \cdot range(x_i - \mu)} + \beta, 
\end{equation}
where $C(B) = 1/{\sqrt{2 \cdot ln(B)}}$ (e.g., 0.32 for $B=128$), $\mu$ is the expectation over $X^{(c)}$, and $range(x) = max(x) - min(x)$.
The constant $C(B)$ is the key to the RN as it helps accurately approximate the standard deviation of $X^{(c)}$.
As feature maps in a DNN model are originated from a sum of many inputs, $X^{(c)}$ naturally follows the Gaussian distribution~\cite{gaussian_distribution}.
Thus, the range of the input $X^{(c)}$ is highly correlated with the standard deviation magnitude.

\subsection{Block Floating Point}\label{sec:BFP}
A floating point number, which is generally used in the DNN training, is represented as
\begin{equation}
    x_i = (-1)^{s_i} \times (1.m_i) \times 2^{e_i},
\end{equation}
where $s_i$ is the sign, $m_i$ is the mantissa, and $e_i$ is the exponent of the number $x_i$.
Block floating point (BFP) is a special form of representing a set of floating point numbers.
In the BFP representation, multiple floating point numbers form a block, say $\vec{x} = [x_1, x_2, ..., x_N]$, that will share an exponent value $e_s$.
Note that the shared exponent is obtained by `$e_s=\lfloor log_2(max(|x_1|,\ldots,|x_N|)) \rfloor$'.
Then, we shift the mantissa of all numbers in the block to the right by ($e_s - e_i$).
As a result, we get a new block of floating point numbers that is 
\begin{equation}
    \vec{x}_{BFP} = [\hat{x_1}, \hat{x_2}, \ldots, \hat{x_N}]\cdot 2^{e_{s}} \approx [x_1, x_2, ..., x_N],
\end{equation}
where $\hat{x_i}$ is the aligned number represented by only `sign + mantissa'.
The main advantage of BFP representation is that it becomes possible to perform a inner product between two floating point vectors
with fixed-point arithmetic units~\cite{hybrid_bfp}.
Another important benefit of utilizing the BFP representation is reducing required memory footprint for storing tensors (thanks to the exponent sharing).
In this work, we focus on the BN, which requires a considerable memory access~\cite{bn_sysml}, thus the purpose of utilizing the BFP format is to minimize the access energy of feature maps during the BN process by reducing the memory footprint.

\section{Motivation}\label{sec:motivation1}
\subsection{Compute Units for BN Layers}\label{sec:compute_unit}
\begin{figure}[t]
    \centering
    \includegraphics[scale=0.58]{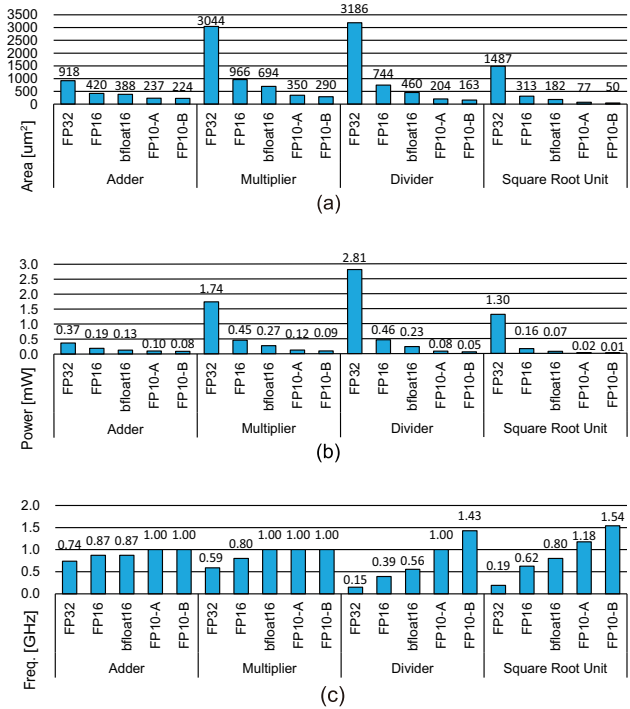}
    \caption{Hardware costs of compute units used in the BN layer are extracted using a 45nm CMOS technology and compared at various FP precisions. The hardware costs are measured by (a) area, (b) power consumption, and (c) maximum operating frequency.}
    \label{fig:compute_unit}
\end{figure}

The prior work on designing an energy-efficient DNN accelerator mostly focus on Conv/FC operations~\cite{rapid, flexblock, cambricon,  bias_training}, 
while there is lack of research on making the BN hardware more efficient.
One of the most effective ways of improving hardware efficiency of a processing unit is reducing the bit-precision. In this work, we extensively study the impact of reduced precision on computations involved in the BN layer.
Prior to analyzing the impact of low-precision BN processing on DNN training accuracy,
we performed a detailed analysis on compute units involved in the BN computation.
There are four main compute units for the BN layer processing: i) adders, ii) multipliers, iii) dividers, and iv) square root units. 
We synthesized these compute units in a 45nm CMOS technology using DesignWare IPs supported by Synopsys~\cite{dw_synopsys}.
Four different FP precisions were tested, i.e., FP32, FP16, bfloat16 and FP10.
Since training DNNs with BN layers computed in FP8 failed, as discussed in Section~\ref{sec:lightnorm}, we report the hardware costs of compute units using FP10 as a lower bound.
Two variants of FP10 are tested since forward and backward passes require different bit configurations to ensure training stability (refer to Section~\ref{sec:Blocked Range Normalization}).
One is FP10-A \{1,5,4\} and another is FP10-B \{1,6,3\}\footnote{The data format is represented by \{sign bit, exponent bits, mantissa bits\} throughout this paper.}.
%and the synthesized results are represented in Fig.~\ref{fig:compute_unit}.
For the fair comparison, the synthesized clock frequency is set to that of FP32 for area and power reports (Fig.~\ref{fig:compute_unit}(a-b)).
As a reference, we also report the maximum clock frequency that each compute unit for a given FP precision can operate at (Fig.~\ref{fig:compute_unit}(c)).
As expected, the lower the FP precision, the lower the occupied area and power consumption.
For instance, we can reduce the area and power consumption by 74.9\% and 75.2\% on average by using FP10 compared to FP32.
By having more exponent bits and less mantissa bits in bfloat16, we can reduce the area and power consumption by 4.8\% and 25.5\% on average compared to FP16.
This is because handling mantissa bits is more complex in floating point arithmetic~\cite{fp4dl}.

\begin{figure}[t]
    \includegraphics[scale=0.52]{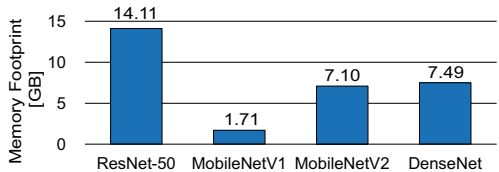}
    \centering
    \caption{The required memory footprints of BN layers per training epoch obtained from four CNN benchmarks.}
    \label{fig:memory_footprint_fig}
\end{figure}

\subsection{Memory Requirement of BN Layers}\label{sec:motivation1}

When training CNNs, feature maps generated at each layer in forward pass are used during backward pass when computing gradients~\cite{bn_sysml}. 
Thus, the feature maps generated in the forward pass need to be stored in DRAM for the later use. 
Fig.~\ref{fig:memory_footprint_fig} shows a required memory capacity per training epoch when various CNN models are trained on CIFAR-100 with mini-batch size of 256.
MobileNetV1 requires the least memory footprint among the tested CNN models. 
This is because MobileNetV1 has the smallest number of layers and reduces the number of computations by using depthwise-separable convolutions. 
On the contrary, ResNet-50 requires the largest memory footprint. 
It requires 2.0$\times$ and 1.8$\times$ more memory capacity than MobileNetV2 and DenseNet-121. 
Since the training is an iterative process, accessing several or tens of GBs of data per epoch will consume significant amount of energy.
Thus, minimizing the data travels from the processing core to DRAM is also an important thing to consider when dealing with BN layers.

\begin{comment}
\begin{figure}[b]
    \includegraphics[scale=0.50]{figures/floating_point_addition_2.eps}
    \centering
    \caption{Algorithm of floating point addition/subtraction} 
    \label{fig:floating point addition} 
\end{figure} 
\end{comment}

\section{LightNorm: Low-Precision Blocked Range Norm}\label{sec:lightnorm}

\subsection{Low-Precision Batch Normalization}\label{sec:low_prec_BN}

To perform BN, we first need to do a channel-wise feature map accumulation for a given mini-batch to compute $\mu$ in Eq. (\ref{eq:bn}) or (\ref{eq:range_norm}).
Prior work on training DNNs perform BN with single precision (FP32) in order to extract the statistics of feature maps as accurate as possible to ensure the training convergence~\cite{mix_training, bfloat16, nnp-t, hybrid_bfp, parameter8_training, minifloat}.
In this section, we analyze the impact of the reduced precision on forward and backward passes at the BN layers to independently set the minimum bit-precision for each pass.

\begin{comment}Fig.~\ref{fig:floating point addition} represents an algorithm of floating point addition/subtraction. 
\end{comment}

\begin{table}[t]
\centering
\caption{Dynamic range and representable value range of various floating point formats. The '$s$', '$e$' or '$m$' represents the length of sign, exponent or mantissa bits.} 
\label{tab:representable_range}
\scalebox{0.8}{
\begin{tabular}{|c|c|c|c|}
\hline
\textbf{Data Format} & \textbf{\{s, e, m\}} & \textbf{Dynamic Range} & \textbf{Representable Range} \\ \hline\hline
FP32     & \{1, 8, 23\} & -126 $\sim$127 & $\pm$ [1.1755E-38, 3.4028E+38] \\ \hline
bfloat16 & \{1, 8, 7\}  & -126 $\sim$127 & $\pm$ [1.1755E-38, 3.3895E+38] \\ \hline
FP16     & \{1, 5, 10\} & -14 $\sim$15   & $\pm$ [6.1035E-05, 6.3488E+04] \\ \hline
FP10-A     & \{1, 5, 4\} & -14 $\sim$15   & $\pm$ [6.1035E-05, 6.3488E+04] \\ \hline
FP10-B     & \{1, 6, 3\} & -30 $\sim$31   & $\pm$ [9.3132E-10, 4.0265E+09] \\ \hline
FP8      & \{1, 5, 2\}  & -14 $\sim$15     & $\pm$ [6.1035E-05, 5.7344E+04] \\ \hline
\end{tabular}%
}
\end{table}

\subsubsection{Floating Point Formats}

In floating point representations, there are three bit components:
a sign bit ($s$), mantissa bits ($m$) and exponent bits ($e$).
Here, the length of mantissa bits determines the precision, and that of exponent bits determines the dynamic range.
Table~\ref{tab:representable_range} summarizes the dynamic range and representable value range of various floating point formats, i.e., FP32, bfloat16, FP16, FP10 and FP8.
The FP32 is the most precise representation with the widest dynamic range among the four.
The bfloat16 is particularly designed for training DNNs with a wider dynamic range compared to FP16, while sacrificing the precision.
Notably, the precision is less important than the dynamic range when considering the DNN training~\cite{bfloat16_tpu, nnp-t}.
%One may utilize a scale factor to shift/bias the representable range so that FP16 can still be used during the DNN training~\cite{mix_training}.
The FP10-A and FP8 provide the same dynamic range with FP16, but with a significantly less precision (merely 4-bit and 2-bit is used as mantissa bits).
With the use FP10-B, we slightly compromise precision for better dynamic range compared to FP10-A.
As seen from Fig.~\ref{fig:compute_unit}, using FP10 results in the most efficient hardware implementation.
Thus, our goal is to make full use of FP10 arithmetic units for BN processing to design an efficient DNN training accelerator without training accuracy loss.

\subsubsection{Proper Length of Exponent Bits}
\begin{figure}[t]
    \includegraphics[scale=0.6]{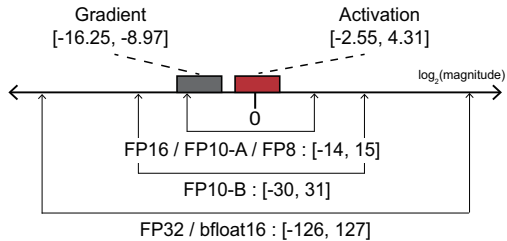}
    \centering
    \caption{Dynamic range of feature maps and gradients in BN layers. It is extracted by training ResNet-50 on CIFAR-100 over 160 epochs.}
    \label{fig:dynamic_range} 
\end{figure}

The dynamic range has an important role in training accuracy. 
This is because values outside the representable range become zero. 
According to~\cite{mix_training}, the omitted values due to the limited dynamic range results in significant training accuracy drop. 
Therefore, we observed data distribution of both forward pass (i.e., activations) and backward pass (i.e., gradients) at BN layers in order to select a proper exponent bits for BN processing. 
Fig.~\ref{fig:dynamic_range} shows the dynamic range of feature maps (activations) and gradients at BN layers of ResNet-50 trained on CIFAR-100 dataset.
The activations have a dynamic range of [-2.55, 4.31], which is safely covered by 5-bit exponent bits (i.e., FP16, FP10-A and FP8). 
%This can be covered by the dynamic ranges of FP16 and FP8. 
However, gradients have a dynamic range of [-16.25, -8.97] which can be covered by 6-bit or higher exponent bits (i.e., FP32, bfloat16 and FP10-B).
According to this data-driven analysis, we decided to select 6-bit or higher exponent bits for BN processing.

\begin{comment}

In order to understand precision requirements of the BN layer, we first extracted the data distributions for both forward and backward passes during the training.
The errors in floating point numbers rise from two sources: i) insufficient mantissa bits and ii) insufficient exponent bits.
The former is critical when many floating point numbers are accumulated together (e.g., computing $\mu$ and $Var[X^{(c)}]$ in Eq.~(\ref{eq:bn})).
When adding two floating point numbers, the mantissa bits of a number with a smaller exponent is shifted to the right, 
which may become zero with insufficient mantissa bits~\cite{fp_adder}.
This phenomenon is referred to as zero setting error (ZSE).
Insufficient exponent bits limit the representable range (Table~\ref{tab:representable_range}) that causes overflow or underflow.
Fig.~\ref{fig:zero_setting_error} shows data distributions of input feature maps at the 3\textsuperscript{rd} layer of ResNet-18
and their normalized features after the BN layer using either FP32 or FP8.
Due to shortened mantissa and exponent bits in FP8, the distribution is distorted with a zero-mean and unit variance, respectively.

The mean and standard deviation of feature maps at various training epochs and data formats are analyzed as shown in Table~\ref{tab:analysis_distribution}.
\end{comment}

\subsubsection{Proper Length of Mantissa Bits} 
Accumulation is a crucial operation to compute layer statistics (e.g., computing $\mu$ and $Var[X^{(c)}]$ in Eq.~(\ref{eq:bn})). 
When performing a floating point addition, the maximum exponent among two numbers is extracted first, then a mantissa of the number with a smaller exponent is shifted to the right by the exponent difference.
If the length of mantissa bits is smaller than the shift amount, the smaller number becomes zero. 
This phenomenon is referred to as zero setting error (ZSE). 
Therefore, insufficient mantissa bits will cause errors when computing $\mu$ and $\sigma$.
The mean and standard deviation of normalized feature maps at various training epochs and data formats are analyzed in Table~\ref{tab:analysis_distribution}.
To look at the sole impact of the FP precision on BN layers during the forward pass, we kept all other computations at FP32.
Due to extremely short mantissa bits in FP8, i.e., 2-bit, the normalized distribution is distorted deviating from zero-mean and unit-variance.
By allowing two additional bits for mantissa, i.e., FP10-A, the computation errors of estimating $\mu$ and $\sigma$ become much smaller than FP8.

\subsubsection{Principle Behind Selection of BN Precision for DNN Training} 

\begin{figure}[t]
    \includegraphics[scale=0.64]{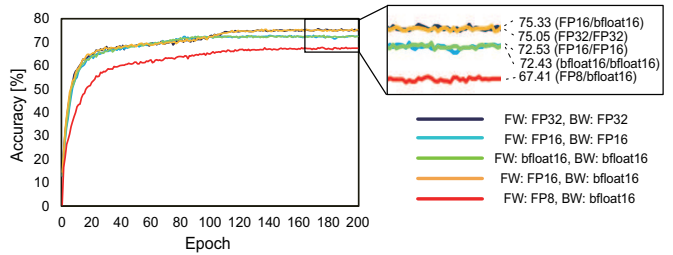}
    \centering
    \caption{Training results of various data formats on ResNet-50 with CIFAR-100 dataset.}
    \label{fig:solution_training_accuracy}
\end{figure}

Based on the analysis so far, we can conjecture that the length of mantissa bits is critical for the forward (FW) pass and that of exponent bits is more critical for the backward (BW) pass at BN layers.
To verify this statement, we conducted a set of experiments on ResNet-50 with CIFAR-100 dataset.
We selected different FP precisions for forward and backward passes to check the sensitivity of the length of mantissa bits or exponent bits at each pass on training accuracy.
The training curves at various FP combinations are provided in Fig.~\ref{fig:solution_training_accuracy}.
The baseline is using FP32 for both forward and backward passes.
When using FP16 or bfloat16 for both forward and backward passes, the training accuracy degrades by $\sim$2.5\%.
Instead, we used FP16 for the forward pass and bfloat16 for the backward pass based on our analysis.
As a result, it was possible to reach similar training accuracy to the FP32 baseline.
However, if we move from FP16 to FP8 for the forward pass while keeping bfloat16 for the backward pass, the training accuracy significantly degrades by 7.9\% due to the insufficient mantissa bits.
In this set of experiments, we used the conventional BN layers.

\begin{table}[t]
\centering
\caption{Mean and standard deviation values of normalized feature maps at the 3\textsuperscript{rd} layer of ResNet-50 at various training epochs with CIFAR-100. The precision of the BN layer for forward pass is only changed while bit widths of all other computations are kept at FP32.}
\label{tab:analysis_distribution}
\scalebox{0.8}{
\begin{tabular}{|c|c|c|c|c|c|c|}
\hline
\multicolumn{1}{|c|}{\textbf{Epoch: 30}} &
  \multicolumn{1}{c|}{\textbf{FP32}} &
  \multicolumn{1}{c|}{\textbf{bfloat16}} &
  \multicolumn{1}{c|}{\textbf{FP16}} &
  \multicolumn{1}{c|}{\textbf{FP10-A}} &
  \multicolumn{1}{c|}{\textbf{FP8}} \\ \hline\hline
\multicolumn{1}{|c|}{\textbf{Mean ($\mu$)}} &
  \multicolumn{1}{c|}{3.9581E-09} &
  \multicolumn{1}{c|}{-2.4620E-06} &
  \multicolumn{1}{c|}{4.1793E-08} &
  \multicolumn{1}{c|}{-0.0003} &
  \multicolumn{1}{c|}{-0.0022} \\ \hline
\multicolumn{1}{|c|}{\textbf{Stdev ($\sigma$)}} &
  \multicolumn{1}{c|}{1.0000} &
  \multicolumn{1}{c|}{1.0002} &
  \multicolumn{1}{c|}{1.0000} &
  \multicolumn{1}{c|}{1.0052} &
  \multicolumn{1}{c|}{1.0150} \\ \hline
\multicolumn{1}{l}{} &
  \multicolumn{1}{l}{} &
  \multicolumn{1}{l}{} &
  \multicolumn{1}{l}{} &
  \multicolumn{1}{l}{} &
  \multicolumn{1}{l}{} \\ \hline
\multicolumn{1}{|c|}{\textbf{Epoch: 50}} &
  \multicolumn{1}{c|}{\textbf{FP32}} &
  \multicolumn{1}{c|}{\textbf{bfloat16}} &
  \multicolumn{1}{c|}{\textbf{FP16}} &
  \multicolumn{1}{c|}{\textbf{FP10-A}} &
  \multicolumn{1}{c|}{\textbf{FP8}} \\ \hline\hline
\multicolumn{1}{|c|}{\textbf{Mean ($\mu$)}} &
  \multicolumn{1}{c|}{1.6298E-09} &
  \multicolumn{1}{c|}{-1.0652E-07} &
  \multicolumn{1}{c|}{3.4925E-09} &
  \multicolumn{1}{c|}{-0.0003} &
  \multicolumn{1}{c|}{-0.0025} \\ \hline
\multicolumn{1}{|c|}{\textbf{Stdev ($\sigma$)}} &
  \multicolumn{1}{c|}{1.0000} &
  \multicolumn{1}{c|}{1.0002} &
  \multicolumn{1}{c|}{1.0000} &
  \multicolumn{1}{c|}{1.0051} &
  \multicolumn{1}{c|}{1.0142} \\ \hline
\multicolumn{1}{l}{} &
  \multicolumn{1}{l}{} &
  \multicolumn{1}{l}{} &
  \multicolumn{1}{l}{} &
  \multicolumn{1}{l}{} &
  \multicolumn{1}{l}{} \\ \hline
\multicolumn{1}{|c|}{\textbf{Epoch: 70}} &
  \multicolumn{1}{c|}{\textbf{FP32}} &
  \multicolumn{1}{c|}{\textbf{bfloat16}} &
  \multicolumn{1}{c|}{\textbf{FP16}} &
  \multicolumn{1}{c|}{\textbf{FP10-A}} &
  \multicolumn{1}{c|}{\textbf{FP8}} \\ \hline\hline
\multicolumn{1}{|c|}{\textbf{Mean ($\mu$)}} &
  \multicolumn{1}{c|}{-1.7462E-09} &
  \multicolumn{1}{c|}{-8.2760E-07} &
  \multicolumn{1}{c|}{-1.6997E-08} &
  \multicolumn{1}{c|}{-0.0003} &
  \multicolumn{1}{c|}{-0.0025} \\ \hline
\multicolumn{1}{|c|}{\textbf{Stdev ($\sigma$)}} &
  \multicolumn{1}{c|}{1.0000} &
  \multicolumn{1}{c|}{1.0002} &
  \multicolumn{1}{c|}{1.0000} &
  \multicolumn{1}{c|}{1.0050} &
  \multicolumn{1}{c|}{1.0147} \\ \hline
\multicolumn{1}{l}{} &
  \multicolumn{1}{l}{} &
  \multicolumn{1}{l}{} &
  \multicolumn{1}{l}{} &
  \multicolumn{1}{l}{} &
  \multicolumn{1}{l}{} \\ \hline
\multicolumn{1}{|c|}{\textbf{Epoch: 100}} &
  \multicolumn{1}{c|}{\textbf{FP32}} &
  \multicolumn{1}{c|}{\textbf{bfloat16}} &
  \multicolumn{1}{c|}{\textbf{FP16}} &
  \multicolumn{1}{c|}{\textbf{FP10-A}} &
  \multicolumn{1}{c|}{\textbf{FP8}} \\ \hline\hline
\multicolumn{1}{|c|}{\textbf{Mean ($\mu$)}} &
  \multicolumn{1}{c|}{1.1642E-09} &
  \multicolumn{1}{c|}{-2.3531E-06} &
  \multicolumn{1}{c|}{3.2596E-08} &
  \multicolumn{1}{c|}{-0.0002} &
  \multicolumn{1}{c|}{-0.0022} \\ \hline
\multicolumn{1}{|c|}{\textbf{Stdev ($\sigma$)}} &
  \multicolumn{1}{c|}{1.0000} &
  \multicolumn{1}{c|}{1.0002} &
  \multicolumn{1}{c|}{1.0000} &
  \multicolumn{1}{c|}{1.0050} &
  \multicolumn{1}{c|}{1.0148} \\ \hline
\end{tabular}%
}
\end{table}

\begin{comment}
\begin{figure}[t]
    \includegraphics[scale=0.25]{figures/zero_setting_error.eps}
    \centering
    \caption{Data distributions of (a) input feature maps and their normalized features after the BN layter computed in (B) FP32 or (c) FP8. The distributions are extracted from $3^{th}$ layer of ResNet-18 at $30^{th}$ epoch when trained on CIFAR-100.}
    \label{fig:zero_setting_error}
\end{figure}  
\end{comment}

\subsection{Blocked Range Normalization}\label{sec:Blocked Range Normalization}
%We found proper precision of each step of BN in previous section. To reduce memory footprint, we adopt block floating point (BFP) representation. Moreover, we replace the BN to RN.

In addition to the reduced precision in the BN layers, we applied two more approximation techniques that mainly focus on reducing
the number of DRAM accesses for higher energy-efficiency.

\subsubsection{Utilizing Range Normalization} 
\begin{figure}[t]
    \includegraphics[scale=0.57]{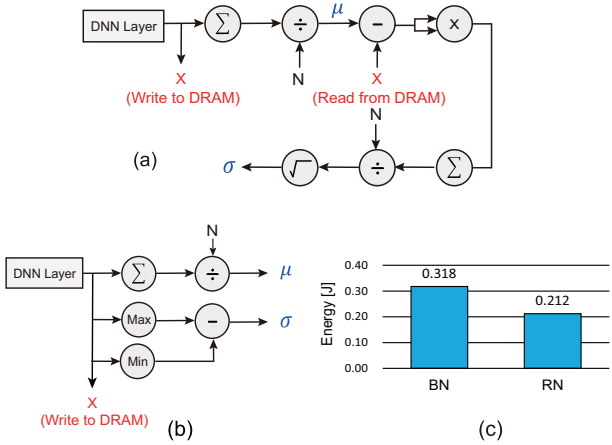}
    \centering
    \caption{Computational flow of (a) the conventional BN and (b) range normalization (RN), and (c) the energy comparison between BN and RN when processing the most memory-intensive BN layer in MobileNetV2.}
    \label{fig:comparison_gn_rn} 
\end{figure}

The range normalization (RN) presented in~\cite{range_bn} has two important advantages which are essential to achieve lightweight BN layers.
First, RN has a fewer number of operations to get layer statistics and significantly less DRAM accesses.
Fig.~\ref{fig:comparison_gn_rn} shows the computational flow to get statistics, i.e., $\mu$ and $\sigma$, in BN and RN.
%The reason why we replace BN to RN is that BN has the fewer number of operations to get statistics and DRAM accesses. Fig.~\ref{fig:comparison_gn_rn} shows the computational flow to get statistics in BN and RN. 
For the conventional BN, standard deviation of X can be calculated after getting $\mu=$ E[X]. 
Thus, feature maps (X) from the previous DNN layer need to be fetched from DRAM to calculate $\sigma$. 
Unlike BN, however, RN can calculate the standard deviation from X directly without the expensive DRAM read, simply by monitoring the max(X) and min(X) for computing the range as the approximate value of $\sigma$. 
%The standard deviation can be calculated while comparing inputs. 
The energy saving by removing the DRAM read was estimated by assuming a 16Gb LPDDR3 module as DRAM~\cite{lpddr3}.
The energy consumption in processing a forward step at a BN layer in MobileNetV2 with CIFAR-100 dataset using BN was 0.318J, while it was 0.212J using RN which translates to 32.7\% energy saving.
%Fig.~\ref{fig:comparison_gn_rn} (c) represents how much energy consumption can be reduced if we use RN instead of BN. The energy was estimated on 16Gb LPDDR3~\cite{lpddr3}, and the layer which requires maximum memory capacity in MobileNetV2 was utilized. 
%To process forward step of BN, 0.069$J$ is required, but only 0.046$J$ is needed in RN. Consequently, 32.7\% energy saving, which is consumed in data fetching from DRAM, is possible when we use RN. Thanks to this attractive point of RN, we exploit RN instead of BN.
Another advantage of RN is that it does not require square ($X^2$) and square root ($\sqrt{Var[X]}$) operations.
The problem with these operations is that they result in intermediate values with a wide dynamic range.
As pointed out previously, if the numbers to be added are widely distributed, a lot of ZSEs occur.
Thus, the conventional BN will require FP precision with sufficient mantissa bits (e.g., FP16).
Since RN does not need these problematic operations, we can safely reduce mantissa bits and ensure stable DNN training.
To verify this, we tried training MobileNets using RN with FP10 formats.
We tried four combinations of two FP10 formats, i.e., FP10-A and FP10-B, as shown in Table~\ref{tab:range_bn_fp10}.
As expected, allowing wider mantissa bits at FW pass and wider exponent bits at BW pass showed similar or slightly higher accuracy compared to the FP32 baseline.
Throughout the remainder of this paper, we use \{1,5,4\} for the FW pass and \{1,6,3\} for the BW pass as the precision for LightNorm (simply denote as `FP10').

%% Let's add a table here to determine the precision with range BN prior to showing the result for blocked range BN
\begin{table}[t]
\centering
\caption{Test accuracy when MobileNets are trained with four different combinations of FP10 formats on CIFAR-100 dataset using range normalization (RN) instead of the conventional BN.} 
\label{tab:range_bn_fp10}
\scalebox{0.8}{
\begin{tabular}{|c|c|c|c|}
\hline
\multicolumn{2}{|c|}{\textbf{Data Format (FW / BW)}} & \textbf{MobileNetV1} & \textbf{MobileNetV2} \\ \hline\hline
\multicolumn{2}{|c|}{\textbf{FP32 / FP32 (with BN; not RN)}}   & \textbf{66.21\%}  & \textbf{65.38\%}  \\ \hline
\multicolumn{2}{|c|}{FP10-A:\{1, 5, 4\} / FP10-A:\{1, 5, 4\}}   & 55.68\%  & 52.36\%  \\ \hline
\multicolumn{2}{|c|}{\textbf{FP10-A:\{1, 5, 4\} / FP10-B:\{1, 6, 3\}}}   & \textbf{67.82\%}  & \textbf{65.73\%}  \\ \hline
\multicolumn{2}{|c|}{FP10-B:\{1, 6, 3\} / FP10-A:\{1, 5, 4\}}   & 54.02\%  & 50.46\%  \\ \hline
\multicolumn{2}{|c|}{FP10-B:\{1, 6, 3\} / FP10-B:\{1, 6, 3\}}   & 66.71\%  & 63.99\%  \\ \hline
\end{tabular}%
}
\end{table}

\subsubsection{Exponent Sharing (Block Floating Point)} 
\begin{figure}
    \includegraphics[scale=0.6]{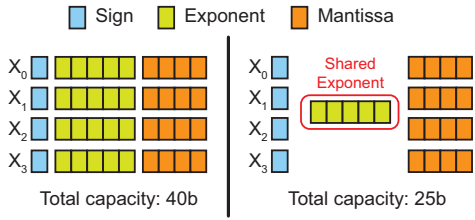}
    \centering
    \caption{Comparison of data size between FP10 and BFP10 formats with the group size of 4.}
    \label{fig:memory_footprint} 
\end{figure}

To further improve the energy efficiency of LightNorm, we group numbers to form a block that shares exponent prior to storing them to DRAM (`Write to DRAM' in Fig.~\ref{fig:comparison_gn_rn}).
This reduces the size of data that move across the expensive DRAM interface.
In addition, reducing the required memory space is important since BN entails large intermediate data as previously shown in Fig.~\ref{fig:memory_footprint_fig}.
This is especially helpful for mobile devices equipped with relatively small DRAM chips.
As an example, assume $N$ floating point numbers are grouped together with group size of $k$, and FP precision used for the BN layer is \{$s,e,m$\}.
Then, the total data size becomes `$N\cdot(s+m)+N/k\cdot e$' instead of `$N\cdot(s+m+e)$'.
Fig.~\ref{fig:memory_footprint} shows a simple example where the tensor size is 4 and FP10-A is used for BN processing.
Without the exponent sharing, the total data size becomes 40-bit.
By representing the tensor in BFP format, i.e., grouping four numbers to have a single shared exponent, the data size reduces to 25-bit ({\it 37.5\% reduction}).
However, if the group size becomes too large, ZSEs will occur for numbers with small exponents.
Note that the shared exponent is the maximum exponent value within the group.
Thus, we need to conservatively set the group size not to hurt the training accuracy.
%The impact of storing intermediate values in the BFP format on training accuracy is provided in Table~\ref{tab:}.

%One of attractive points of BFP is that required memory capacity is reduced. Since BN entails huge memory capacity which comes from intermediate results of BN~\cite{bn_sysml}, reducing memory capacity is important. In floating pint number system, a tensor takes its size as $O(ks+kn+km)$ where k is a tensor size, s, n and m are lengths of sign, mantissa and exponent, respectively. However, in BFP arithmetic, the cost is reduced to $O(ks+kn+m)$, because only one exponent exists within the tensor. Fig.~\ref{fig:memory_footprint} represents an example when the tensor size is 4, and bit-width of data is 10-bit. The floating point number case requires 40-bit of memory capacity in order to store 4 elements, but only 25-bit is required in the case of BFP representation. Meanwhile, we found proper bit-width of each step of BN. We change the formats to BFP representation in order to reduce memory footprint. In other words, FP10-A and FP10-B are replaced to BFP10-A and BFP10-B, respectively. They show low accuracy drop compared to original formats that we found. The impacts of BFP formats on DNN training are represented in section~\ref{sec:lightnorm_acc}.
%\jaeha{Need more detailed explanation on how much memory footprint can be reduced by using the BFP format? Maybe providing a simple equation on the reduction factor on memory will be great.}.

%Meanwhile, although the process to get a mean is identical, the process of RN to get a standard deviation is far simpler. Furthermore, computing time is reduced about twice. It is because only one accumulation over all elements of mini-batch, which is dominant in computation time, is required in RN, whereas BN needs two times of this work. Thanks to these attractive points of RN, we exploit RN instead of BN.
%\seockhwan{modified, 05/21 by Seock-Hwan}

% 여기서부터 수정 필요..
\subsection{Training Accuracy with LightNorm}\label{sec:lightnorm_acc}
%\subsubsection{Methodology}\label{methodology1}
We fuse three techniques explained in Section~\ref{sec:low_prec_BN} and \ref{sec:Blocked Range Normalization} to realize LightNorm, i.e., an extremely cost-effective BN layer.
To evaluate the training accuracy of LightNorm, we replaced the conventional BN layers to LightNorm layers in PyTorch framework. 
We added new classes, i.e., `lightnorm.nn.BatchNorm2d' and `lightnorm.nn.LayerNorm', and used instead of `torch.nn.BatchNorm2d' and `torch.nn.LayerNorm'.
To configure LightNorm layers in PyTorch, we provide a configuration file that has the group size for the exponent sharing and the precision level (FP10 as a default).
To verify the sole impact of LightNorm, we kept other layers such as Conv and FC layers in FP32.
For the evaluation, we selected four CNN benchmarks, i.e., ResNet-50, MobileNetV1, MobileNetV2 and DenseNet-121, trained on CIFAR-100 dataset~\cite{cifar}.
All training hyperparameters are kept the same as training the networks in FP32.
Table~\ref{tab:training_result} summarizes the test accuracy of CNN benchmarks using LightNorm trained with various group sizes.
We tested the group size of 4, 8 or 16 to group feature maps or gradients during the FW and BW passes.
As shown in Table~\ref{tab:training_result}, allowing group size of 8 or larger results in a large amount of ZSEs significantly degrading the training accuracy.
With the group size of four the test accuracy is similar to the FP32 baseline.
The test accuracy only drops by 0.5\% on average.
Therefore, for the design of LightNorm hardware in the following section, 
we use BFP10 (FW: \{1,5,4\}, BW: \{1,6,3\}) as the BN precision using blocked range BN with the group size of 4.
%Furthermore, LightNorm trainer can get a configuration file with respect to the block size from user. 
%Training work can be performed on the size, if the block size is provided. As for precision, a single precision is exploited only except LightNorm; in other words, DNN layers and non-linear functions are performed on a FP32 format in LightNorm trainer. 
%In the forward step of LightNorm, BFP representation of FP8 is utilized, but accumulation is performed by increasing mantissa length as 8-bit. 
%In the backward step of LightNorm, BFP16 is utilized. 
%We report impacts of LightNorm on our target networks, which are obtained by LightNorm trainer. 
%As for dataset, CIFAR-100~\cite{cifar} was exploited. 

%\subsubsection{Training Results of BRN}\label{training_results_brn}

\begin{table}
\centering
\caption{Test accuracy of four CNN benchmarks using LightNorm when trained with various group sizes (4, 8 and 16).}
\label{tab:training_result}
\scalebox{0.8}{%
\begin{tabular}{|c|cccc|}
\hline
\textbf{Network} &
  \multicolumn{1}{c|}{\textbf{ResNet-50}} &
  \multicolumn{1}{c|}{\textbf{MobileNetV1}} &
  \multicolumn{1}{c|}{\textbf{MobileNetV2}} &
  \textbf{DenseNet-121} \\ \hline\hline
\textbf{FP32} & 
  \multicolumn{1}{c|}{\textbf{75.71\%}} & 
  \multicolumn{1}{c|}{\textbf{66.21\%}} & 
  \multicolumn{1}{c|}{\textbf{65.38\%}} & 
  \multicolumn{1}{c|}{\textbf{75.14\%}} \\ \hline
\textbf{BFP10, group=4} & 
  \multicolumn{1}{c|}{\textbf{73.68\%}} & 
  \multicolumn{1}{c|}{\textbf{67.71\%}} & 
  \multicolumn{1}{c|}{\textbf{64.84\%}} & 
  \multicolumn{1}{c|}{\textbf{74.20\%}} \\ \hline
\textbf{BFP10, group=8} & 
  \multicolumn{1}{c|}{71.95\%} & 
  \multicolumn{1}{c|}{38.48\%} & 
  \multicolumn{1}{c|}{28.65\%} & 
  \multicolumn{1}{c|}{73.43\%}  \\ \hline
\textbf{BFP10, group=16} & 
  \multicolumn{1}{c|}{55.10\%} & 
  \multicolumn{1}{c|}{18.21\%} & 
  \multicolumn{1}{c|}{18.93\%} & 
  \multicolumn{1}{c|}{63.58\%}  \\ \hline
\end{tabular}%
}
\end{table}

\section{DNN Training Accelerator with LightNorm}\label{sec:accelerator}

\subsection{LightNorm Hardware}\label{sec:lightnorm_hw}

\begin{figure}[b]
    \includegraphics[scale=0.34]{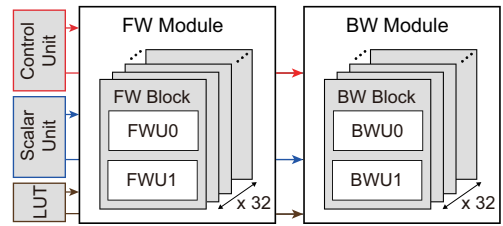}
    \centering
    \caption{Overall architecture of LightNorm hardware.}
    \label{fig:BlockNorm_Archtiecture_fig} 
\end{figure}

To support the end-to-end training, DNN training accelerators should be equipped with BN hardware.
Fig.~\ref{fig:BlockNorm_Archtiecture_fig} shows the overall architecture of LightNorm hardware.
It consists of one forward (FW) pass module and backward (BW) pass module to support hardware-accelerated training of BN layers, 
which processes 32 channels in parallel (directly connected to columns of a systolic array).
Note that the systolic array is typically used at accelerating general matrix multiply (GEMM) operations~\cite{bfloat16_tpu}.
In addition, LightNorm hardware has a control unit, a scalar unit and a lookup table (LUT).
The scalar unit calculates `-$\frac{\gamma}{\sigma+\epsilon}$' and `$(\sigma)^{-3/2} \cdot \frac{\gamma C(B)}{2}$' that are required in the backward pass of LightNorm.
The LUT stores a pre-computed $C(B) = 1/ \sqrt{2 \cdot ln(B)}$ for various $B$ values, where $B$ is the mini-batch size.
In our design, it stores $C(B)$ values when $B=$ 16, 32, 64, 128, 256 and 1024.
Note that LightNorm follows the RN computation given by Eq.~(\ref{eq:range_norm}) and channel-wise normalization is performed for a given mini-batch.

\subsubsection{Forward Pass Module (LightNorm - FW)}\label{sec:FW_Core}

\begin{figure}[t]
    \includegraphics[scale=0.50]{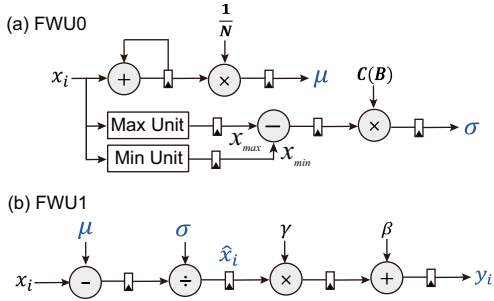}
    \centering
    \caption{The structure of a forward pass module using FP10-A: (a) FWU0 for computing $\mu$ and $\sigma$, and (b) FWU1 for normalizing $x_i$.}
    \label{fig:fwu0} 
\end{figure}

The forward pass module is dedicated to a forward pass of the LightNorm layer, which uses `FP10-A' format.
It has 32 FW blocks for parallel execution of BN for 32 output channels.
The outputs from the training accelerator, i.e., a 32$\times$32 systolic array in this paper, are streamed into 32 FW blocks.
Each FW block has two forward pass units that are FWU0 and FWU1 (Fig.~\ref{fig:fwu0}).
The streamed $x_i$'s in FWU0 are accumulated by FP10-A adder to compute $\mu$ of a particular output channel.
At the same time, Max and Min units extract the maximum and minimum values of $x_i$ for `$\sigma = C(B)\cdot(x_{max}-x_{min})$' computation.
Then, computed $\mu$ and $\sigma$ are passed to FWU1 for the actual normalization on $x_i$.
These two units are pipelined, which means that FWU0 takes inputs every clock cycle and FWU1 normalizes the feature map by using the pre-computed $\mu$ and $\sigma$.

%% From here, need to change, 05/31
\subsubsection{Backward Pass Module (LightNorm - BW)}\label{sec:BW_Core}

\begin{figure}[t]
    \includegraphics[scale=0.50]{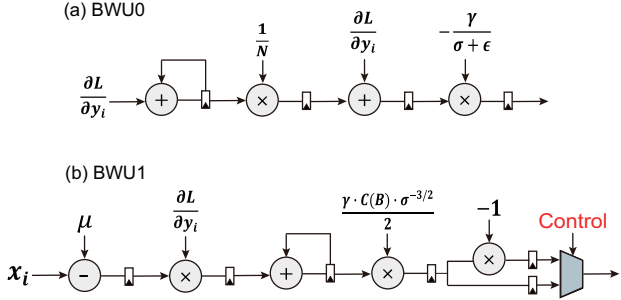}
    \centering
    \caption{The structure of a backward pass module using FP10-B: (a) BWU0 for backpropagating errors through the numerator of Eq.~(\ref{eq:range_norm}), and (b) BWU1 for backpropagating errors through the denominator of Eq.~(\ref{eq:range_norm}).}
    \label{fig:bwu0} 
\end{figure}

The backward pass module backpropagates local gradients through the LightNorm layer, which uses `FP10-B' format. 
Similar to the FW module, it consists of 32 BW blocks where each block consists of two backward pass units that are BWU0 and BWU1 (Fig.~\ref{fig:bwu0}).
The BWU0 computes the local gradient $({\partial L}/{\partial x_i})_1$ through the numerator of Eq.~(\ref{eq:range_norm}), which can be defined as
%Meanwhile, equations of gradients are different depending on $x_i$ in LightNorm. The equation of gradients, which are not $x_{max}$ and $x_{min}$, is as follows.
\begin{equation}\label{eq:rn_bw_eq0}
    (\frac{\partial L}{\partial x_i})_1 = -\frac{\gamma}{\sigma+\epsilon}\cdot(\frac{1}{N}\cdot\Sigma^{N}_{i=1}\frac{\partial L}{\partial y_i}+\frac{\partial L}{\partial y_i}),
\end{equation}
where ${\partial L}/{\partial y_i}$ is the gradient arrived at the output of a LightNorm layer, $N$ is the number of elements in local gradients per channel, and $\gamma$ is the coefficient of the BN layer, $\sigma$ is the standard deviation, and $\epsilon$ is used to ensure stability.
%BWU0 processes above equation. 
Another local gradient $({\partial L}/{\partial x_i})_2$ through the denominator of Eq.~(\ref{eq:range_norm}), i.e., $range(x_i-\mu)$ function, can be obtained by 
 %\begin{equation}
 %   -\frac{\gamma}{nm\sigma^2}\cdot\frac{\partial L}{\partial y_i}
 %\end{equation}
 \begin{equation}\label{eq:rn_bw_eq1}
    (\frac{\partial L}{\partial x_i})_2 =
    \frac{\gamma\cdot C(B)}{2}\cdot(\sigma)^{-3/2}\cdot \{\Sigma^{n}_{i=1}\frac{\partial L}{\partial y_i}\cdot (x_i - \mu)\}.
 \end{equation}
If $x_i=x_{min}$, the final local gradient at input ($x_i$) of the LightNorm layer is computed by $({\partial L}/{\partial x_i})_1+({\partial L}/{\partial x_i})_2$.
If $x_i=x_{max}$, the final local gradient at $x_i$ is computed by $({\partial L}/{\partial x_i})_1-({\partial L}/{\partial x_i})_2$.
Otherwise, the local gradient at $x_i$ simply equals to $({\partial L}/{\partial x_i})_1$.
Since BWU1 computes Eq.~(\ref{eq:rn_bw_eq1}), we need to selectively provide a positive or negative value of $({\partial L}/{\partial x_i})_2$.
This is controlled by the multiplexer placed inside the BWU1 unit.

\subsection{Evaluation of LightNorm Hardware}\label{sec:eval_lightnorm}
\subsubsection{Methodology and Baselines}\label{sec:eval_method} 

In order to evaluate energy efficiency of LightNorm hardware, we implemented RTL and synthesized it at 150MHz with Synopsys Design Compiler using 45nm open cell library~\cite{si2_open_lib}. 
To compare LightNorm with other baselines, we designed BN hardware for the conventional BN~\cite{batchnorm} and restructured BN~\cite{bn_sysml} with FP32 compute units.
Then, they are synthesized at the same clock frequency using the same technology node.
In the conventional BN, variance of a tensor $X$ is computed by
\begin{equation}\label{eq:conventional_bn}
    Var[X] = E[X-E[X]]^2,
\end{equation}
where $X$ is the mini-batch of input tensor per channel.
In the conventional BN, the variance can be calculated only after computing the mean $E[X]$. 
This temporal dependency results in two DRAM accesses for fetching the entire $X$'s for a given mini-batch size ({\it bandwidth-limited}).
%Therefore, two times of memory sweepings over inputs are required: One is for the mean, the other is for the variance. 
To reduce the overhead of excessive DRAM accesses, the restructured BN calculates the variance in a different manner, which is
\begin{equation}\label{eq:restructured_bn}
    Var[X] = E(X^2) - E(X)^2.
\end{equation}
By simple restructuring of Eq.~(\ref{eq:conventional_bn}) to (\ref{eq:restructured_bn}),
the restructured BN computes the mean and variance in parallel. 
This effectively reduces the number of DRAM accesses by half.
Meanwhile, local gradients at the conventional and restructured BN layers are calculated by the same equation~\cite{bn_sysml},
which is
 \begin{equation}
 \label{equation_backward_bn}
    \frac{\partial L}{\partial x_i} = \frac{\gamma}{\sqrt{\sigma^2+\epsilon}}\cdot(\frac{\partial L}{\partial y_i}-\frac{1}{N}\cdot\frac{\partial L}{\partial \beta}-\frac{1}{N}\cdot\frac{(x_i - \mu)}{\sqrt{\sigma^2+\epsilon}}\cdot\frac{\partial L}{\partial \gamma}),
\end{equation}
where $\frac{\partial L}{\partial \gamma} = \Sigma^{N}_{i=1}\frac{\partial L}{\partial y_i}\cdot\frac{(x_i - \mu)}{\sqrt{\sigma^2+\epsilon}}$ and $\frac{\partial L}{\partial \beta} = \Sigma^{N}_{i=1}\frac{\partial L}{\partial y_i}$.
For the performance comparison, a cycle-approximate simulator was designed, which outputs the estimated clock cycles for three BN hardware modules (i.e., LightNorm and two baselines).
The hardware evaluation was performed on four benchmarks, i.e., ResNet-50, MobileNetV1, MobileNetV2 and DenseNet-121, using CIFAR-100 dataset.

\begin{comment}
The simulator outputs number of cycles, by considering memory accesses as well as computation time on compute units. Note that we used CACTI~\cite{cacti, cactip} and parameters of DRAM from~\cite{micron_parameter} of DDR4-2666 in order to get energy consumption in system level. Based on the power and cycle information, we extracted energy-efficiency information of three accelerators.
\end{comment}

\subsubsection{Area and Power Consumption}\label{area_power_result}

\begin{table}[t]
\centering
\caption{Area and power breakdowns of LightNorm hardware.}
\label{tab:area_power}
\scalebox{0.78}{%
\begin{tabular}{|c|c|c|}
\hline
\textbf{Module}                       & \textbf{Area [$\mu$m\textsuperscript{2}]} & \textbf{Power [mW]} \\ \hline\hline
\textbf{LightNorm - FW}            & 68961.68 (74.62\%)              & 2.1344 (53.32\%)               \\ \hline
\textbf{LightNorm - BW}            & 22115.31 (23.93\%)              & 1.8200 (45.37\%)               \\ \hline
\textbf{Scalar Unit}           & 1011.42 (1.09\%)              & 0.0515 (1.28\%)               \\ \hline
\textbf{Others}        & 327.16 (0.35\%)              & 0.0056 (0.14\%)               \\ \hline\hline
\textbf{Total}         & 92415.57 (100.00\%)              & 4.0115 (100.00\%)                \\ \hline
\end{tabular}%
}
\end{table}

\begin{comment}
\begin{figure}[b]
    \includegraphics[scale=0.88]{figures/area_power2.eps}
    \centering
    \caption{Breakdowns of areas (a) and powers (b) of LightNorm and comparison targets}
    \label{fig:area_power} 
\end{figure}
\end{comment}

Table~\ref{tab:area_power} shows area and power breakdowns of the LightNorm hardware. 
It reports the area and power consumption of all modules which are shown in Fig.~\ref{fig:BlockNorm_Archtiecture_fig}. 
For the LUT and control unit, the area and power consumption are reported together as `Others'. 
In total, LightNorm hardware occupies about 0.09mm$^2$ of area and consumes 4.01mW of power.
The areas of the (conventional) BN and restructured BN are 1.44mm$^2$ and 1.54mm$^2$, respectively.
The power consumptions of the BN and restructured BN are 59.61mW and 63.53mW, respectively.
For the precision independent comparison, we also designed the LightNorm hardware with FP32 compute units.
Then, it occupies 0.99mm$^2$ of area and consumes 37.04mW of power.
Thus, 1.5$\times$ smaller area and 1.7$\times$ lower power consumption on average are used by LightNorm hardware even with the same precision (FP32), {\it thanks to the use of range BN}.
With the use of FP10-A for the forward pass and FP10-B for the backward pass, LightNorm takes up 16.2$\times$ smaller area and consumes 15.4$\times$ lower power on average.
This benefit comes from the {\it fused approximation schemes, i.e., reduced precision, range BN, and exponent sharing}, presented in Section~\ref{sec:lightnorm}.

\begin{comment}
Note that BN calculates a standard deviation with $\sqrt{E[X-E[X]]^2}$, and restructured BN computes it with $\sqrt{E(X^2) - E(X)^2}$.
\end{comment}

\subsubsection{Cycle Estimation}\label{sec:perf_lightnorm}

\begin{figure}[t]
    \includegraphics[scale=0.46]{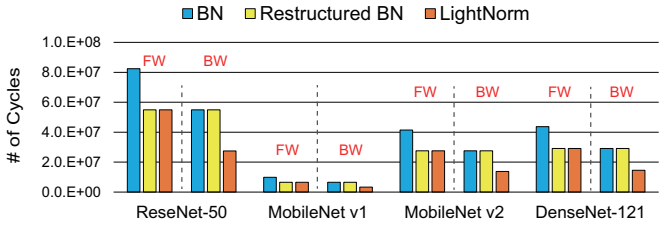}
    \centering
    \caption{Comparison in terms of required clock cycles to compute the BN layers on both forward (FW) and backward (BW) passes. The BN processing with three different BN hardware modules are compared, i.e., (conventional) BN, restructured BN, and LightNorm.} 
    \label{fig:cycle_bn}
\end{figure}

Fig.~\ref{fig:cycle_bn} shows the estimated clock cycles of the BN layers in four representative benchmarks with a mini-batch size of 256.
The BN processing with the (conventional) BN, restructured BN, and LightNorm hardware are compared. 
For the forward pass (FW), the restructured BN consumes 33.3\% less clock cycles than the conventional BN on average. 
This is because the restructured BN calculates $\mu$ and $\sigma$ in parallel, while the BN computes $\mu$ and $\sigma$ sequentially. 
However, they consume a similar number of cycles in the backward pass (BW), since they perform exactly the same operations in the BW pass. 
For all benchmarks, LightNorm significantly reduces the required clock cycles by 1.5$\times$ and 2.0$\times$ on average compared to conventional BN for the FW and BW passes, respectively.
%compared to the BN and restructured BN, respectively.
As emphasized in the previous sections, LightNorm achieves this speed-up by fusing three approximation techniques, i.e., reduced precision, exponent sharing, and range BN.
Thanks to these approximation techniques, operations in the both FW and BW passes are simplified.
This is obvious by comparing Eq.~(\ref{eq:rn_bw_eq0}) to Eq.~(\ref{equation_backward_bn}).

\subsubsection{Energy Analysis}\label{sec:energy_lightnorm}

\begin{comment}
\begin{figure}[t]
    \includegraphics[scale=0.44]{figures/energy_comparision.eps}
    \centering
    \caption{Comparison of energy consumption on target networks.}
    \label{fig:energy_analysis}
\end{figure}

\end{comment}

\begin{comment}
Fig.~\ref{fig:energy_analysis} shows energy consumption of our LightNorm hardware and comparison targets (i.e., conventional BN and restructured BN) which are implemented by BFP8+8 and FP32, respectively. Furthermore, it also represents LightNorm hardware designed with a high precision (i.e., FP32 data format). The hardware based on RN has largest power consumption, but it has lower energy consumption compared to conventional BN. This comes from reduced latency in forward pass of normalization. The LightNorm hardware, which is implemented by FP32, has fewer energy consumption than comparison targets. Replacement of BN to RN results in far lower energy consumption than comparison targets. The LightNorm with BFP8+8, finally, achieves lowest energy consumption. It has $13.32\times$ and $11.13\times$ lower average energy consumption than conventional BN and restructured BN cases. 
\end{comment}

%The conventional BN and restructured BN, which utilize a single precision, have energy consumption to process one epoch as 14.8mJ and 12.4mJ, respectively. 
The energy consumptions of the conventional BN and restructured BN with FP32 are 14.8mJ and 12.4mJ on average to perform one training epoch.
Although the hardware based on restructured BN has larger power consumption than the conventional BN case (Section~\ref{area_power_result}), it consumes less energy. 
This comes from the reduced clock cycle in the forward pass. 
LightNorm implemented with FP32 has 2.8$\times$ and 2.4$\times$ consumes lower energy than the conventional BN and restructured BN, respectively. 
Finally, LightNorm with FP10 only consumes 0.6mJ of energy on average.
%has very low energy consumption. This takes only 0.6mJ average energy. 
Owing to the fused approximation schemes (i.e., reduced precision, exponent sharing and range BN), LightNorm consumes 23.5$\times$ and 19.6$\times$ less energy than the conventional BN and restructured BN, respectively.

%% Commented Out
\begin{comment}
\begin{figure}[h]
    \includegraphics[scale=0.56]{figures/energy_efficiency_final.eps}
    \centering
    \caption{Average energy-cost on target networks to process one mini-batch}
    \label{fig:energy_efficiency}
\end{figure}

Fig.~\ref{fig:energy_efficiency} represents average energy consumption of LightNorm and comparison targets on four target networks to process one mini-batch size of 128. The LightNorm, implemented by BFP16+8, has 7.80$\times$ and 7.71$\times$ smaller energy consumptions than BN and restructured BN. Moreover, LightNorm has 3.20$\times$ smaller energy consumption than BlockNorm implemented with a single precision. 
\end{comment}
%%%%

\begin{comment}
\subsubsection{Memory Footprint}\label{sec:footprint_lightnorm} 
Since fetching data from DRAM consumes significant amount of energy~\cite{eyeriss}, we need to reduce the amount of data movement between the training accelerator and off-chip memory. 
Due to the low data locality of the BN process~\cite{bn_sysml},
it becomes the memory-intensive operation.
By reducing the precision to FP10 in LightNorm instead of the widely used FP32 in the prior work, the required data movement reduces by 70.0\% on average.
By storing the feature maps (FW) or local gradients (BW) in the BFP format,
we can further reduce the required memory footprint by 15.1\% reaching 80.6\% reduction in total by using LightNorm.
\end{comment}

\subsection{Training Accelerator with LightNorm}\label{sec:end_to_end_trainer}

\subsubsection{Methodology}\label{methodology3}

\begin{figure}[t]
    \includegraphics[scale=0.35]{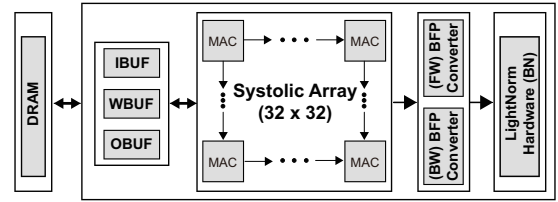}
    \centering
    \caption{Overall architecture of a DNN training accelerator equipped with LightNorm hardware.}
    \label{fig:training_accelerator}
\end{figure}

To look at practical effectiveness of LightNorm, we designed a training accelerator that consists of data buffers, a 32$\times$32 systolic array, and LightNorm hardware with BFP converters (Fig.~\ref{fig:training_accelerator}).
%Fig.~\ref{fig:training_accelerator} represents an overall architecture of the accelerator. 
%It is composed of TPU-like systolic array (SA) and BlockNorm in order to process DNN layer and LightNorm. 
To perform a system-level analysis on the training accelerator, we also considered DRAM and SRAM accesses when estimating the performance and energy consumption. 
As for DRAM, 16Gb LPDDR3 was assumed and its associated timing specifications are used~\cite{lpddr3}.
For the power and timing analysis of SRAM blocks, we used CACTI-6.0~\cite{cacti6}.
The sizes of on-chip buffers are selected differently depending on the precision levels used by various training accelerators as summarized in Table~\ref{tab:design_configuration}.
When deciding the on-chip buffer size, we also considered hiding the DRAM access latency to keep the systolic array busy as much as possible.
To estimate the clock cycle consumed by the systolic array, we modified an open-source cycle-level simulator, i.e., Scale-Sim~\cite{scalesim}, to consider our SRAM and DRAM configurations. 
The number of clock cycles consumed by the LightNorm hardware is measured by the RTL simulation by assuming ImageNet-scale images~\cite{imagenet} as inputs to the network for a more realistic analysis.
The RTLs of all hardware configurations for the evaluation are synthesized at 150MHz with Synopsys Design Compiler using 45nm CMOS technology.

\begin{table*}[t]
\centering
\caption{Various hardware design configurations of a DNN training accelerator for the area and energy analysis.}
\label{tab:design_configuration}
\scalebox{0.74}{%
\begin{tabular}{|c|c|c|c|c|c|c|c|}
\hline
\textbf{Hardware Configurations}              & \textbf{HW1} & \textbf{HW2} & \textbf{HW3} & \textbf{HW4} & \textbf{HW5} & \textbf{HW6} & \textbf{HW7 (Proposed)} \\ \hline\hline
\textbf{Precision in SA (Mul. / Add.)}     & FP32 / FP32     & FP32 / FP32     & FP32 / FP32     & FP8 / FP32    & FP8 / FP32    & FP8 / FP32    & FP8 / FP32   \\ \hline
\textbf{Precision in BN (FW / BW)} & FP32 / FP32       & FP32 / FP32      & FP32 / FP32       & bfloat16 / bfloat16       & bfloat16 / bfloat16       & bfloat16 / bfloat16       & BFP10 / BFP10    \\ \hline
\textbf{Batch Norm Type}         & Conventional BN         & Restructured Norm         & Range Norm (RN)  & Conventional BN         & Restructured Norm         & Range Norm (RN)  & LightNorm  \\ \hline
\textbf{Bus Size (I / W / O) {[}bit{]}}    & 1024 / 1024 / 1024 & 1024 / 1024 / 1024 & 1024 / 1024 / 1024 & 256 / 256 / 1024 & 256 / 256 / 1024 & 256 / 256 / 1024 & 256 / 256 / 256 \\ \hline
\textbf{Memory sizes (I / W / O) {[}KB{]}} & 128 / 128 / 64    & 128 / 128 / 64    & 128 / 128 / 64    & 32 / 32 / 32  & 32 / 32 / 32  & 32 / 32 / 32  & 32 / 32 / 24    \\ \hline
\end{tabular}
}%
\end{table*}

\begin{figure*}[t]
    \includegraphics[scale=0.43]{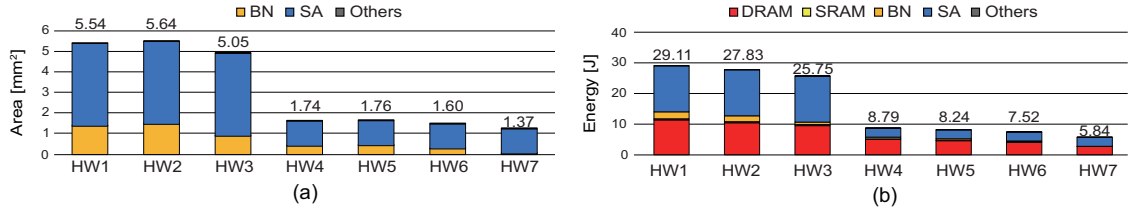}
    \centering
    \caption{(a) Area and (b) energy breakdowns of various DNN training accelerators. High-performance training accelerators (HW1$\sim$3), energy-efficient training accelerators (HW4$\sim$6), and the proposed training accelerator are compared.}
    \label{fig:training_accelerator_area_power}
\end{figure*}

\subsubsection{Hardware Configurations}\label{training_accelerators}

Fig.~\ref{fig:training_accelerator} shows the overall architecture of a training accelerator equipped with LightNorm hardware. 
For the comparison, there are various hardware configurations that can affect the area and power consumption of the training accelerator.
One configurable parameter is the bit-precision of the systolic array (SA).
The systolic array processes Conv/FC layers and it has 32$\times$32 multiply-accumulate (MAC) units.
The precision for the multiplier at each MAC unit can be changed, i.e., FP32 for the high-performance training accelerators~\cite{mix_training,batchnorm} and FP8 for the energy-efficient training accelerators~\cite{rapid,8bit_training}.
Not to lose the training accuracy, we use the FP32 adder at each MAC unit for all hardware configurations.
Another configurable parameter is the type of BN layer and its precision level (e.g., BFP10 for LightNorm).
In Table~\ref{tab:design_configuration}, HW1$\sim$3 are high-performance training accelerators with different types of BN layers using FP32.
HW4$\sim$6 are energy-efficient training accelerators with different BN modules using bfloat16, which is based on the recent studies possibly enabling on-device training~\cite{8bit_training, parameter8_training, bfloat16}.
There are three different BN types compared in this work, i.e., conventional BN~\cite{batchnorm}, restructured BN~\cite{bn_sysml}, and range BN~\cite{range_bn}.
The BFP converter in Fig.~\ref{fig:training_accelerator} becomes simply a quantization unit for HW4$\sim$6 changing FP32 outputs to bfloat16 values for more efficient BN processing.
HW7 is the proposed training accelerator using low-precision systolic array with LightNorm hardware.
Note that BFP10 is used at the BN layers in LightNorm and BFP converters are used to quantize outputs to FP10 and store them in a more memory-efficient way, i.e., BFP10, as presented in Section~\ref{sec:lightnorm}.

\subsubsection{Area Analysis}\label{area_trainers}

As expected, high-performance training accelerators, i.e., HW1$\sim$3, which use FP32 MAC units and FP32 BN modules occupy larger area (Fig.~\ref{fig:training_accelerator_area_power}(a)).
For instance, HW1 takes up 3.2$\times$ larger area than HW4 even though both use the same BN type. 
Despite of the identical precision level in HW1$\sim$3, different BN types lead to different chip areas. 
For example, HW3 using RN occupies 10.4\% smaller area than HW2 using restructured BN. 
Since HW4$\sim$6 use lower precision in both SA and BN modules, they occupy 68.6\% smaller area on average than the high-performance training accelerators (HW1$\sim$3). 
Note that the area of the BN module slightly reduces, this is because all adders in the BN hardware are still performed with FP32 as in~\cite{bfloat16}. 
The proposed training accelerator (HW7) occupies the smallest area thanks to the use of low-precision SA and LightNorm hardware.
The area saving in the BN hardware comes from the reduced precision and more hardware-friendly RN computations.
It occupies 4.0$\times$ and 1.2$\times$ smaller areas on average than HW1$\sim$HW3 and HW4$\sim$HW6, respectively.

\subsubsection{Energy Analysis}\label{energy_trainers}

The analysis on energy consumption of each training accelerator is performed on one training epoch.
%Different BN hardwares lead to differences in energy consumption. 
Among HW1$\sim$3, a RN-based accelerator (HW3) shows the best energy-efficiency (11.6\% better than HW1). 
HW4$\sim$6 show 3.4$\times$ lower energy consumption on average than HW1$\sim$3. 
This reduction is mainly due to i) low-precision multipliers in both SA and BN, and ii) reduction in the data access energy.
%Although they have similar energy consumption on BN modules, i) utilizing low bit-precision compute unit in SA (i.e., 8-bit multiplier) and ii) storing data with reduced precision bring about reduced energy. 
The proposed accelerator (HW7) shows the minimum energy consumption, i.e., 1.3$\sim$5.0$\times$ lower energy than other accelerators.
Compared to HW1$\sim$3 and HW4$\sim$6, HW7 saves the energy by 78.8\% and 28.6\%, respectively.
This is due to the proposed LightNorm as BN processing, which minimizes both BN processing energy and memory access energy.

\section{Conclusion}

In recent DNNs, the relative importance in the execution time and energy consumption of the batch normalization (BN) process has been significantly increased.
In this paper, therefore, we presented an extremely memory- and energy-efficient BN process, named LightNorm. 
To achieve this goal, we fused three approximation techniques, which are i) low bit-precision, ii) range batch normalization, and iii) block floating point. 
%We analyzed data distribution of BN layers and found proper precisions which are robust to ill effects of low bit-precision in BN such as zero setting error and unrepresentable range.
These techniques are carefully selected that help reducing the complexity of BN layers and improving its hardware efficiency without sacrificing the DNN training accuracy.
To demonstrate the hardware efficiency, we designed a customized LightNorm hardware and compared with the other conventional BN hardware designs.
Finally, we extended the hardware evaluation by integrating the LightNorm hardware to a real DNN training accelerator. 
In conclusion, LightNorm has improved the energy-efficiency by 1.3$\sim$5.0$\times$ compared to various configurations of DNN training accelerators.

\begin{comment}

\section*{Acknowledgements}
This document is derived from previous conferences, in particular ICCD 2022.
\end{comment}

%%%%%%% -- PAPER CONTENT ENDS -- %%%%%%%%

%%%%%%%%% -- BIB STYLE AND FILE -- %%%%%%%%
\bibliographystyle{IEEEtranS}
\bibliography{refs}

% Generated by IEEEtranS.bst, version: 1.14 (2015/08/26)
\begin{thebibliography}{10}
\providecommand{\url}[1]{#1}
\csname url@samestyle\endcsname
\providecommand{\newblock}{\relax}
\providecommand{\bibinfo}[2]{#2}
\providecommand{\BIBentrySTDinterwordspacing}{\spaceskip=0pt\relax}
\providecommand{\BIBentryALTinterwordstretchfactor}{4}
\providecommand{\BIBentryALTinterwordspacing}{\spaceskip=\fontdimen2\font plus
\BIBentryALTinterwordstretchfactor\fontdimen3\font minus
  \fontdimen4\font\relax}
\providecommand{\BIBforeignlanguage}[2]{{%
\expandafter\ifx\csname l@#1\endcsname\relax
\typeout{** WARNING: IEEEtranS.bst: No hyphenation pattern has been}%
\typeout{** loaded for the language `#1'. Using the pattern for}%
\typeout{** the default language instead.}%
\else
\language=\csname l@#1\endcsname
\fi
#2}}
\providecommand{\BIBdecl}{\relax}
\BIBdecl

\bibitem{range_bn}
R.~Banner \emph{et~al.}, ``Scalable methods for 8-bit training of neural
  networks,'' in \emph{Proc. of NeurIPS}, 2018.

\bibitem{understanding_batchnorm}
N.~Bjorck \emph{et~al.}, ``Understanding batch normalization,'' in \emph{Proc.
  of NeurIPS}, 2018.

\bibitem{gpt3}
T.~Brown \emph{et~al.}, ``Language models are few-shot learners,'' in
  \emph{Proc. of NeurIPS}, 2020.

\bibitem{parameter8_training}
L.~Cambier \emph{et~al.}, ``Shifted and squeezed 8-bit floating point format
  for low-precision training of deep neural networks,'' in \emph{Proc. of
  ICLR}, 2020.

\bibitem{batchnorm_residual_block}
S.~De \emph{et~al.}, ``Batch normalization biases residual blocks towards the
  identity function in deep networks,'' in \emph{Proc. of NeurIPS}, 2020.

\bibitem{imagenet}
J.~Deng \emph{et~al.}, ``Imagenet: A large-scale hierarchical image database,''
  in \emph{Proc. of CVPR}, 2009.

\bibitem{bert}
J.~Devlin \emph{et~al.}, ``{BERT:} pre-training of deep bidirectional
  transformers for language understanding,'' \emph{arXiv:1810.04805}, 2018.

\bibitem{hybrid_bfp}
M.~Drumond \emph{et~al.}, ``Training {DNNs} with hybrid block floating point,''
  in \emph{Proc. of NeurIPS}, 2018.

\bibitem{minifloat}
S.~Fox \emph{et~al.}, ``A block minifloat representation for training deep
  neural networks,'' in \emph{Proc. of ICLR}, 2021.

\bibitem{mit_book}
I.~Goodfellow \emph{et~al.}, \emph{Deep Learning}.\hskip 1em plus 0.5em minus
  0.4em\relax MIT Press, 2016.

\bibitem{bfloat16_tpu}
{Google Cloud}, ``{BFloat16}: The secret to high performance on cloud {TPUs},''
  2019.

\bibitem{resnet}
K.~He \emph{et~al.}, ``Deep residual learning for image recognition,'' in
  \emph{Proc. of CVPR}, 2016.

\bibitem{nnp-t}
B.~Hickmann \emph{et~al.}, ``Intel {Nervana} neural network processor-t
  {(NNP-T)} fused floating point many-term dot product,'' in \emph{Proc. of
  ARITH}, 2020.

\bibitem{norm_matter}
E.~Hoffer \emph{et~al.}, ``Norm matters: efficient and accurate normalization
  schemes in deep networks,'' in \emph{Proc. of NeurIPS}, 2018.

\bibitem{densenet}
G.~Huang \emph{et~al.}, ``Convolutional networks with dense connectivity,''
  \emph{IEEE TPAMI}, 2019.

\bibitem{batchnorm}
S.~Ioffe \emph{et~al.}, ``Batch normalization: Accelerating deep network
  training by reducing internal covariate shift,'' in \emph{Proc. of ICML},
  2015.

\bibitem{fp4dl}
J.~Johnson, ``Rethinking floating point for deep learning,''
  \emph{arXiv:1811.01721}, 2018.

\bibitem{bn_sysml}
W.~Jung \emph{et~al.}, ``Restructuring batch normalization to accelerate {CNN}
  training,'' in \emph{Proc. of SysML}, 2019.

\bibitem{bfloat16}
D.~Kalamkar \emph{et~al.}, ``A study of {BFLOAT16} for deep learning
  training,'' \emph{arXiv:1905.12322}, 2019.

\bibitem{cifar}
A.~Krizhevsky \emph{et~al.}, ``{CIFAR}-10 and {CIFAR}-100 dataset,'' 2010.

\bibitem{driving}
S.~Lin \emph{et~al.}, ``The architectural implications of autonomous driving:
  Constraints and acceleration,'' in \emph{Proc. of ASPLOS}, 2018.

\bibitem{mix_training}
P.~Micikevicius \emph{et~al.}, ``Mixed precision training,'' in \emph{Proc. of
  ICLR}, 2018.

\bibitem{lpddr3}
{Micron}, ``{Mobile LPDDR3 SDRAM: 178-Ball, Single-Channel Mobile LPDDR3 SDRAM
  Features},'' \url{https://www.micron.com/products/dram/lpdram/16Gb}, 2014.

\bibitem{cacti6}
{N. Muralimanohar}, ``{CACTI 6.0: A Tool to Model Large Caches},''
  \url{https://www.hpl.hp.com/techreports/2009/HPL-2009-85.pdf}, 2009.

\bibitem{blur}
Nimisha \emph{et~al.}, ``Blur-invariant deep learning for blind-deblurring,''
  in \emph{Proc. of ICCV}, 2017.

\bibitem{flexblock}
S.-H. Noh \emph{et~al.}, ``{FlexBlock}: A flexible {DNN} training accelerator
  with multi-mode block floating point support,'' \emph{arXiv:2203.06673},
  2022.

\bibitem{bias_training}
J.~Park. \emph{et~al.}, ``A {40nm 4.81TFLOPS/W 8b} floating-point training
  processor for non-sparse neural networks using shared exponent bias and
  24-way fused multiply-add tree,'' in \emph{Proc. of ISSCC}, 2021.

\bibitem{scalesim}
A.~Samajdar \emph{et~al.}, ``{SCALE-Sim}: Systolic cnn accelerator simulator,''
  \emph{arXiv:1811.02883}, 2018.

\bibitem{mobilenet_v2}
M.~Sandler \emph{et~al.}, ``{MobileNetV2}: Inverted residuals and linear
  bottlenecks,'' in \emph{Proc. of CVPR}, 2018.

\bibitem{batchnorm_mit}
S.~Santurkar \emph{et~al.}, ``How does batch normalization help optimization?''
  in \emph{Proc. of NeurIPS}, 2018.

\bibitem{si2_open_lib}
{Si2}, ``{15nm Open-cell Library and 45nm FreePDK},''
  \url{https://si2.org/open-cell-library/}, 2022.

\bibitem{gaussian_distribution}
D.~Soudry \emph{et~al.}, ``Expectation backpropagation: Parameter-free training
  of multilayer neural networks with continuous or discrete weights,'' in
  \emph{Proc. of NeurIPS}, 2014.

\bibitem{dw_synopsys}
{Synopsys}, ``{DesignWare IP},''
  \url{https://www.synopsys.com/designware-ip.html}, 2022.

\bibitem{rapid}
S.~Venkataramani \emph{et~al.}, ``{RaPiD: AI} accelerator for ultra-low
  precision training and inference,'' in \emph{Proc. of ISCA}, 2021.

\bibitem{8bit_training}
N.~Wang \emph{et~al.}, ``Training deep neural networks with 8-bit floating
  point numbers,'' in \emph{Proc. of NeurIPS}, 2018.

\bibitem{cambricon}
Y.~Zhao \emph{et~al.}, ``{Cambricon-Q}: A hybrid architecture for efficient
  training,'' in \emph{Proc. of ISCA}, 2021.

\end{thebibliography}
%%%%%%%%%%%%%%%%%%%%%%%%%%%%%%%%%%%%

\end{document}